# Efficient model for low-energy transverse beam dynamics in a nine-cell 1.3 GHz cavity

Thorsten Hellert,[*] Martin Dohlus, and Winfried Decking

*DESY, Notkestrasse 85, 22603 Hamburg, Germany*

(Received 22 May 2017; published 12 October 2017)

FLASH and the European XFEL are SASE-FEL user facilities, at which superconducting TESLA cavities are operated in a pulsed mode to accelerate long bunch-trains. Several cavities are powered by one klystron. While the low-level rf system is able to stabilize the vector sum of the accelerating gradient of one rf station sufficiently, the rf parameters of individual cavities vary within the bunch-train. In correlation with misalignments, intrabunch-train trajectory variations are induced. An efficient model is developed to describe the effect at low beam energy, using numerically adjusted transfer matrices and discrete coupler kick coefficients, respectively. Comparison with start-to-end tracking and dedicated experiments at the FLASH injector will be shown. The short computation time of the derived model allows for comprehensive numerical studies on the impact of misalignments and variable rf parameters on the transverse intra-bunch-train beam stability at the injector module. Results from both, statistical multibunch performance studies and the deduction of misalignments from multibunch experiments are presented.

DOI: 10.1103/PhysRevAccelBeams.20.100702

## I. INTRODUCTION

At FLASH (Free-Electron Laser in Hamburg) [1,2] and European XFEL (European X-Ray Free-Electron Laser) [3,4], superconducting 9-cell TESLA (TeV-Energy Superconducting Linear Accelerator) cavities [5] accelerate the electron bunches in pulsed operation. Due to the high achievable duty cycle, thus long radio-frequency (rf) pulse structure, bunch-trains containing up to 800 and 27000 bunches can be provided at FLASH and European XFEL, respectively. Several cavities with individual operational limits [6] are supplied by one rf power source. Within the bunch-train, the low-level-rf system [7,8] is able to restrict the variation of the vector sum of the accelerating gradient of one rf station sufficiently [9]. However, individual cavities have an intrinsic variation of rf parameters within one bunch-train, caused by the effects of beam loading and Lorentz force detuning [10]. Misaligned cavities in combination with variable rf parameters induce intrabunch-train trajectory variations which decrease the multibunch FEL performance significantly [10].

Understanding the complexity of intrabunch-train trajectory variations is therefore vitally important for a successful multibunch FEL operation at FLASH and European XFEL. In this paper we focus on the transverse dynamics in the injector module. Considering the low beam energy, thus beam sensitivity to off-axis fields, the injector module is of great importance in limiting intrabunch-train trajectory variations. There are several ways for getting a proper description of the transverse beam dynamics in a rf accelerating structure, e.g., using tracking algorithms [11] or simplified analytic models [12,13]. Assuming knowledge of the electromagnetic field distribution and the initial conditions of the particles, tracking provides accurate solutions, even for very low particle energies. Since the track step has to be small compared to a cell length, many steps are required, which needs considerable computation time for simulations with high dimensional parameter scans. Furthermore most tracking codes are not optimized for dealing with different bunches with different rf parameters. Established simplified analytic models on the other hand may calculate the beam transport by few matrix multiplications. However, they are based on assumptions, most importantly ultrarelativistic beams, which do not apply at most particle injectors. Thus, a major challenge is to set up a model for low particle energies $\gamma = [10\ldots 200]$, which is simple enough in order to calculate its output within milliseconds, yet able to reproduce key features of rf dynamics such as rf focussing and coupler kicks. Our approach uses a combination of numerically calculated axially symmetrical beam transport matrices and discretized coupler kicks, coefficients of which are derived via a Runge-Kutta tracking algorithm using a high precision 3D field map of the TESLA cavity. The final model uses the matrix formalism to calculate the beam transport through an accelerating module consisting out of eight cavities in the order of ms for 400 bunches. Its output will be benchmarked against start-to-end tracking and experimental data. The short computation time of the model function allows for comprehensive numerical studies on the impact of misaligned accelerating structures and rf parameters on the intrabunch-train transverse dynamics in the injector module. It will be shown that a distinct determination of the misalignments from fitting experimental data is prevented by experimental constraints. However, the model will be used in order to give statistically meaningful predictions about the multibunch performance of the injector module regarding differential limits of rf parameters and misalignments.

[*]thorsten.hellert@desy.de

## II. MODEL DEVELOPMENT

Figure 1 shows a schematic drawing of the TESLA cavity. It is a 9-cell standing wave structure of about 1 m length whose lowest TM mode resonates at 1.3 GHz. Two field maps are available for describing the electromagnetic field configuration.

The cavity fundamental mode is nominally described by its axial electric field $E_z(z) = E_z(r = 0, z)$. The corresponding field map is obtained from a 2D $[r, z]$ simulation of the TESLA cavity and can be found in Ref. [14]. The field configuration $\mathbf{E}(r \neq 0, z)$ follows from the Maxwell equations. The bold letter indicates a vector. This axially symmetrical (RZ) field map describes the accelerating mode without geometric disturbances.

The 3D field map [15] describes this mode including the fields induced by both higher order mode (HOM) and power coupler and can be found in Ref. [16]. It is given as a table of sine and cosine like amplitudes, $\mathbf{E}_r^{\cos}(\mathbf{r})$ and $\mathbf{E}_r^{\sin}(\mathbf{r})$, respectively, for a purely reflected wave, thus a wave traveling from the cavity into the waveguide and $\mathbf{r} = [x, y, z]$. The electric field of the reflected wave, $\mathbf{E}_r(\mathbf{r}, t)$, has the following time-space-dependency

$$\mathbf{E}_r(\mathbf{r}, t) = \mathbf{E}_r^{\cos}(\mathbf{r})\cos(\omega t) + \mathbf{E}_r^{\sin}(\mathbf{r})\sin(\omega t), \quad (1)$$

with $\omega$ being the angular frequency. Using the Maxwell equations, the electric filed for the forward wave $\mathbf{E}_f(\mathbf{r}, t)$ can be calculated by reversing time and follows as

$$\mathbf{E}_f(\mathbf{r}, t) = \mathbf{E}_r^{\cos}(\mathbf{r})\cos(\omega t) - \mathbf{E}_r^{\sin}(\mathbf{r})\sin(\omega t). \quad (2)$$

Let $A_{[f/r]}$ and $\phi_{[f/r]}$ being the amplitude and phase of the forward and reflected wave to/from the power coupler, respectively. The overall electric field component for the general case with given accelerating voltage $V_0$ and phase $\phi$ with respect to the beam can then be calculated with

$$\mathbf{E} = \Re[V_0/\bar{V}_r e^{i(\omega t + \phi)} \cdot (\mathbf{E}_r^{\cos} + i\Gamma \cdot \mathbf{E}_r^{\sin})] \quad (3)$$

from the 3D field map for the pure decay mode, thus no incoming wave. $\bar{V}_r$ normalizes the field to the Eigenmode-solution of the field map. The voltage standing wave ratio

$$\Gamma = (A_r e^{i\phi_r} - A_f e^{i\phi_f})/(A_r e^{i\phi_r} + A_f e^{i\phi_f}) \quad (4)$$

describes the ratio between the difference of the forwarded and reflected wave in respect to the overall accelerating field. The magnetic component behaves analogously, using similar symmetry properties of the field components. Please note that $\Gamma$ is defined at planes in the waveguide at which the standing wave mode has maximum field amplitude, e.g., at the end of the coupler antenna. Note also that beam loading induces a reflected wave which is not included in Eq. (3).

### A. Beam transport in axially symmetrical cavities

Bunches are described as single particles, hence only referring to their centroid dynamics. Space charge effects and intra bunch wakefields are not considered. The change of transverse coordinates of a particle induced by an axially symmetrical cavity can be written in terms of a matrix formalism as

$$\mathbf{u} = M_{\text{RZ}} \cdot \mathbf{u}_0, \quad (5)$$

with $\mathbf{u}$ and $\mathbf{u}_0$ holding the particle transverse input and output coordinates $\mathbf{u} = [x, x', y, y']$, respectively, and $M_{\text{RZ}}$ being the beam transport matrix of the cavity. The derivation of the analytic beam transport matrix in Ref. [12] assumes the ultrarelativistic limit. The typical initial beam energy in the injector module is about 5 MeV. At this energy the solution of the beam transport equation (5) using the analytical matrix shows significant disagreement with tracking, as can be seen on the left-hand side of Fig. 2. Plotted is the difference of horizontal offset $\Delta x$ in respect to an ASTRA [11] tracking for different initial beam energies. The accelerating gradient is 20 MV m$^{-1}$ For each energy 5000 randomly distributed beam initial trajectories in the range $\mathbf{x}_0 = \pm[4~\text{mm}, 4~\text{mrad}]$ are evaluated. In the energy range of the injector module the disagreement is of up to almost 500 $\mu$m at the end of one cavity. Taking into account the beam size of about 1 mm this is not acceptable for modeling the data. In order to rely on the matrix formalism for describing the beam transport, numerical adjustments on the transfer matrix have to be made.

Using the Maxwell equations, a *quasi*-3D field map can be calculated from Ref. [14]. A Runge-Kutta algorithm is used to solve the equation of motion for one cavity for an ensemble of initial particles, entering the cavity at different offsets and angles. The calculation of the beam transport matrix then becomes a linear regression problem of the form

$$\arg\min_{M_{RZ}} \left( \sum_i |M_{RZ} \cdot \mathbf{u}_{0,i} - \mathbf{u}_{1,i}| \right) \quad (6)$$

where $\mathbf{u}_{0/1,i}$ are the transverse coordinates of the beam at the entrance and at the end of the cavity. However, explicit information of the dependencies of the matrix elements on the initial beam energy $E_0$, the amplitude $V_0$ and phase $\phi$ of the accelerating field is lost in the numerical process. It is therefore necessary to calculate the numerical beam transport matrix at a sufficiently fine grid of model parameters. A cubic spline interpolation of each matrix element is used to obtain the transfer matrix at each point in the parameter space $[E_0 \times V_0 \times \phi]$. The variation of the matrix elements with respect to the model parameters is moderate and smooth. The interpolated solution converges reasonably fast with increased grid points. For this work $14 \times 7 \times 11 = 1309$ grid points within $E_0 = [5…150]$ MeV, $V_0 = [13…30]$ MV m$^{-1}$ and $\phi = \pm 30°$ are chosen. The right-hand side of Fig. 2 shows the difference between the solution of the beam transport equation (5) using the numerical transfer matrix and the previously described ASTRA reference. The agreement in the energy range of the injector module is in the order of 0.1 $\mu$m. It is well below the required accuracy, considering the corresponding beam size of about 1 mm. Therefore, the transverse dynamics related to axially symmetrical rf fields can reasonably well be described by the numerically derived beam transport matrix. The energy gain $\Delta E$ of a particle in the TESLA cavity is determined by the accelerating mode and is to a very good approximation independent of the coupler fields. It's dependency on the beam energy $E_0$, accelerating phase $\phi$ and gradient $V_0$ is found to be well described via

$$\Delta E = \left( a_1 - \frac{a_2 \sin(\phi + a_3)}{E_0 - a_4} \right) \cdot V_0 \cdot \cos(\phi). \quad (7)$$

for the previously mentioned parameter range with fitted coefficients $a_i$.

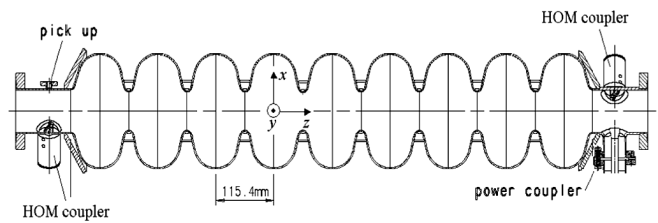

FIG. 1. Longitudinal cross-section of a TESLA cavity [5].

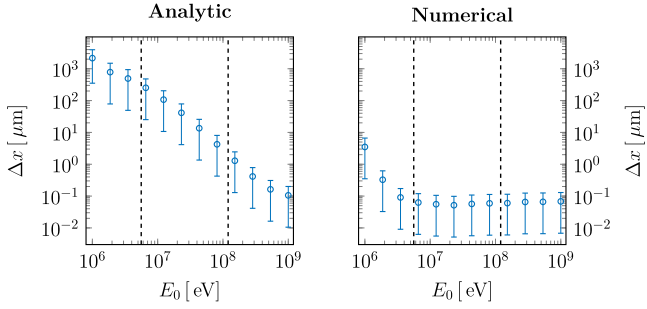

FIG. 2. Comparison between ASTRA tracking and beam transport calculated with analytically (left) and numerically (right) derived transfer matrices. The difference of horizontal offset $\Delta x$ in respect to an ASTRA reference is plotted as a function of initial beam energy. The range between the dashed lines indicates the range of the typical beam energy in the injector module at both, FLASH and E-XFEL.

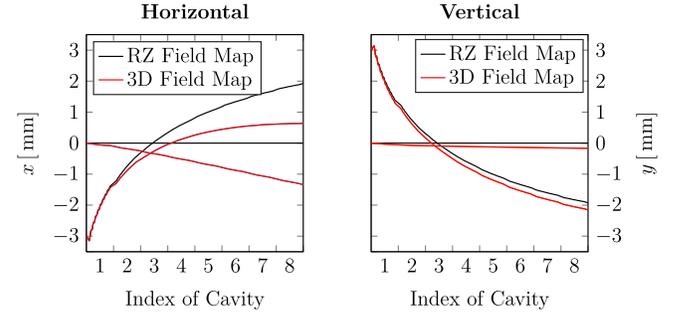

FIG. 3. Particle trajectories through eight cavities for two different initial conditions, plotted for the horizontal (left) and vertical (right) plane. The red lines show the solution of the equation of motion using the 3D field map, the black lines correspond to the solution obtained via the axially symmetrical (RZ) field map. The initial beam parameters are the same for both maps. The initial beam energy is $E_0 = 5$ MeV. Note the strong cavity focusing in the first cavity.

### B. Discrete coupler kicks

HOM and power coupler break the axial symmetry of the cavity and influence the transverse beam dynamics considerably, as illustrated in Fig. 3. Plotted are two particle trajectories through the injector module, where for both trajectories the equation of motion is solved independently for both, the RZ and the 3D field map. The difference between the trajectories obtained with different field maps sums up significantly, especially on the horizontal plane. An appropriate description of the transverse motion therefore has to incorporate the field disturbances caused by the couplers. The transverse rf kick is the total beam transverse momentum change along the trajectory normalized by the longitudinal momentum of the beam. The integrated transverse field strength experienced by an ultrarelativistic paraxial particle

$$\mathbf{V}_\perp(x,y) = \int dz [\mathbf{E}_\perp + c\mathbf{e}_z \mathbf{B}] e^{i\frac{\omega z}{c}}, \quad (8)$$

does not only depend on the absolute distance from axis, $r$, but from $x$ and $y$ independently if the rotational symmetry is broken by couplers. The integrated transverse field induced by the couplers

$$\mathbf{V}_{\text{coupler}}(x,y) = \mathbf{V}_\perp(x,y) - \mathbf{V}_{RZ}(r) \quad (9)$$

can be separated from the axially symmetrical rf focussing part of the field, $\mathbf{V}_{RZ}$, using the 3D field map of the TESLA cavity and extracting the monopole part. The real part of $V_{\text{coupler}}$ corresponds to a net deflection of the bunch centroid, whereas the imaginary part represents a kick which depends on the phase between the oscillating field and each particle. This time-dependent kick induced by the couplers distorts the longitudinal slices of the beam by a different amount, which results in an increase of the projected emittance [17]. The normalized kick $\mathbf{k} = [x', y']$ on a bunch centroid induced by a coupler can be calculated as [18]

$$\mathbf{k}(x,y) = \frac{eV_0}{E_0} \Re\{\tilde{\mathbf{V}}_\perp(x,y) \cdot \exp^{i\phi}\} \quad (10)$$

with $E_0$ being the beam energy and $x$ and $y$ the beam's transverse position at the coupler position. $V_0$ is the amplitude of the accelerating field and $e$ the electron charge. $\tilde{\mathbf{V}}_\perp$ is the normalized complex kick factor, defined as

$$\tilde{\mathbf{V}}_\perp(x,y) = \frac{\mathbf{V}_{\text{coupler}}(x,y)}{\mathbf{V}_\parallel} \quad (11)$$

with $\mathbf{V}_\parallel = \int dz\, \mathbf{e}_z \mathbf{E}(0,0,z)\, e^{\frac{i\omega z}{c}}$. $\tilde{\mathbf{V}}_\perp$ holds the information of the axially asymmetrical field disturbances induced by the couplers. Its real part is plotted in Fig. 4 for both, the upstream and the downstream coupler. Different $\Gamma$, thus modes of cavity operation are evaluated. The kick induced by the upstream HOM coupler does not depend on $\Gamma$. This is due to the fact that the electromagnetic field away from the power input coupler is to a very good approximation described by a standing wave and is not affected by the ratio of the forward and reflected traveling wave. The static part of the downstream kick in respect to different $\Gamma$ relates to the downstream HOM coupler. The $\Gamma$-dependent part relates to the power coupler, which primarily acts horizontally. Coupler kicks can therefore respond independently from the resonating accelerating field to variations of the forward and reflected traveling wave.

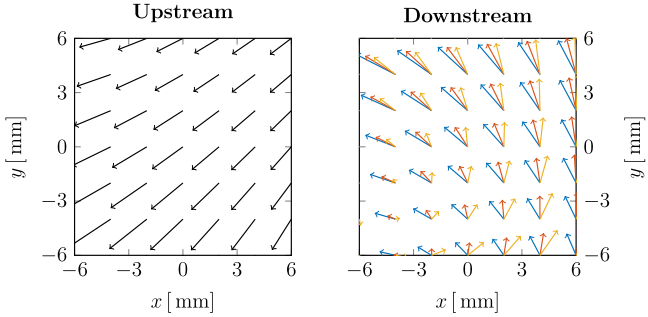

FIG. 4. Real part of the normalized complex kick factor for the upstream (left) and downstream couplers (right) as a function of the transverse coordinates $x$ and $y$. All vectors are scaled by the same amount in order to assure a quantitative comparison. The three colors in the right correspond to the case of pure filling (blue) of the cavity, e.g., no reflected wave and $\Gamma = -1$, standing-wave operation (red, $\Gamma = 0$) and pure decay mode (yellow, $\Gamma = 1$), where there is no incoming wave. Note that the kick induced by the downstream HOM coupler does not change for different $\Gamma$. The net effect of the power coupler is primarily horizontal.

In order to describe the transverse beam dynamics properly we use the axial-symmetric beam transport matrices and insert linearized discrete kicks at a certain location. The kick $\mathbf{k}$ on a bunch's centroid induced by a coupler can therefore be expressed as

$$\mathbf{k}(x,y) \approx \frac{eV_0}{E_0} \cdot \left[ \begin{pmatrix} V_{0x} \\ V_{0y} \end{pmatrix} + \begin{pmatrix} V_{xx} & V_{xy} \\ V_{yx} & V_{yy} \end{pmatrix} \cdot \begin{pmatrix} x \\ y \end{pmatrix} \right]. \quad (12)$$

The normalized complex kick coefficients $V_{ij}$ describe the normalized transverse deflection induced by the coupler and have to be found numerically, since the paraxial assumption of Eq. (8) is not fulfilled at low beam energy. The full beam transport equation of one cavity becomes

$$\mathbf{u}_1 = M_{\text{down}}^{\text{RZ}} \cdot \mathbf{k}_{\text{down}}(M_{\text{center}}^{\text{RZ}} \cdot \mathbf{k}_{\text{up}}(M_{\text{up}}^{\text{RZ}} \cdot \mathbf{u}_0)) \quad (13)$$

where $M_i^{\text{RZ}}$ are the axially symmetrical beam transport matrices calculated according to Eq. (6) between the corresponding reference planes. $M_{\text{up}}^{\text{RZ}}$ describes the beam transport between the entrance of the cavity and the first coupler, $M_{\text{center}}^{\text{RZ}}$ between the upstream and downstream coupler and $M_{\text{down}}^{\text{RZ}}$ between the downstream coupler and the exit of the cavity. $\mathbf{k}_{\text{up}}(\mathbf{u})$ and $\mathbf{k}_{\text{down}}(\mathbf{u})$ evaluate the normalized upstream and downstream coupler kick, respectively, at the transverse coordinate $\mathbf{u} = [x, x', y, y']$, such that $\mathbf{k}(\mathbf{u}) = [x, x' + k_x(x,y), y, y' + k_y(x,y)]$.

For low beam energy the $V_{ij}$ depend not only on the mode of cavity operation, thus the real and imaginary part of the voltage standing wave ratio $\Gamma$, but also implicitly on the particle's initial energy $E_0$ and on both amplitude $V_0$ and phase $\phi$ of the accelerating gradient, spanning a parameter space $E_0 \times V_0 \times \phi \times \Re\Gamma \times \Im\Gamma$. The reason for this is the beam trajectory dependence on these parameters, since the ultrarelativistic limit is not reached. The parameter fit is done as follows: At every point in this parameter space a particle's centroid distribution is created at the entrance of the cavity. The particle distribution at the exit of the cavity is obtained via tracking using the 3D-field map. In addition, the particle distribution in the center of the cavity is recorded. This gives three reference distributions, before and after each coupler region, for which the coupler induced field disturbances are taken into account. Next, the axially symmetrical beam transport matrices are determined: the tracking is redone with the same initial particle distribution and the same rf parameters using the RZ-field map. This time, the particle distribution is recorded additionally at the coupler positions. Between each of these 5 points the axially symmetrical beam transport matrices are calculated according to Eq. (6). For both couplers the reference distributions are compared with the output calculated with the linear beam transport using Eq. (13). A fitting routine was used to then find the $V_{ij}$ which best describe the coupler kick of Eq. (12) using the ultra-relativistic limit [18] as a starting point. The global variation of the $V_{ij}$ were found to be described via

$$V_{ij} = a_1 V_0 \cdot \Re\Gamma \cdot \Im\Gamma + a_2 V_0 \cdot \Re\Gamma + a_3 V_0 \cdot \Im\Gamma \\ + a_4 \Re\Gamma \cdot \Im\Gamma + a_5 V_0 + a_6 \cdot \Re\Gamma + a_7 \Im\Gamma + a_8 \quad (14)$$

with

$$a_n(E_0, \phi) = A_n(E_0) \cdot \cos\phi + B_n(E_0) \cdot \sin\phi + C_n(E_0),$$

$$A_n(E_0) = \frac{y_n^A E_0 + z_n^A}{E_0 - w}, \quad B_n(E_0) = \frac{y_n^B E_0 + z_n^B}{E_0 - w}$$

$$C_n(E_0) = \frac{y_n^C E_0 + z_n^C}{E_0 - w} \quad (15)$$

The $[w, y_n^A, z_n^A, y_n^B, z_n^B, y_n^C, z_n^C]_{ij}$ are 49 constants for each coefficient $V_{ij}$ and were found with a fitting routine.

## III. MODEL VALIDATION

The developed model is compared to the results of a start-to-end tracking using ASTRA [11] and to experimentally derived data at FLASH. The evaluation limits are $E_0 = [5…150]$ MeV, $V_0 = [13…30]$ MV m$^{-1}$, $\phi = \pm 30°$, $\Gamma_{\Re,\Im} = \pm 3$, $u_0 = \pm 6$ mm, $u_0' = \pm 6$ mrad.

### A. Comparison with ASTRA

The rf and beam input parameters are randomly created within the limits. The rms difference of the transverse position $u$ for one cavity as a function of beam input energy is shown in Fig. 5 using different models for calculating the beam transport. At energies above 100 MeV the ultrarelativistic limit is by a very good approximation reached and the beam transport can be calculated according to Ref. [12] including coupler kicks according to Ref. [18]. Especially in the first cavities, however, it is important to use the fitted solutions for both the transfer matrices and the coupler kick coefficients $V_{ij}$. Finally the beam transport through the whole injector module including eight cavities is studied. The initial beam energy is set to $E_0 = 5.6$ MeV. Rf- and the beam input parameter are randomly created within the range of modeling as defined above. The rms difference of the output of the model function using the previously mentioned high energy approach is $\langle\Delta u_{\mathrm{rms}}\rangle = 3$ mm, the mixed approach results in $\langle\Delta u_{\mathrm{rms}}\rangle = 1.3$ mm. The difference of the low energy approach to the ASTRA reference is $\langle\Delta u_{\mathrm{rms}}\rangle = 56$ μm. Taking the beam size of about 1 mm into account this is a satisfactory result.

### B. Comparison with dedicated experiments

This section gives a comparison of the model function with dedicated experiments at the injector section at FLASH. The experimental setup is illustrated in Fig. 6. Two beam position monitors (BPM) are located in the drift space between the solenoid magnet of the rf gun and the injector module. Therefore, the beam input parameter $\mathbf{u}_0 = [x_0, x_0', y_0, y_0']$ at $z_0 = 1.31$ m can be calculated. The measured position at the BPM downstream the last cavity gives the reference $\mathbf{u}_{\mathrm{ref}} = [x_{\mathrm{ref}}, y_{\mathrm{ref}}]$ at $z_{\mathrm{ref}} = 13.43$ m. The model function is evaluated for each bunch independently. The rf signals of the forward and reflected traveling wave are measured inside the waveguides at the circulators. The rf parameters required for the model in order to calculate the beam transport matrices and coupler kick coefficients are $[V_{\mathrm{refl}}, \phi_{\mathrm{refl}}, V_{\mathrm{forw}}, \phi_{\mathrm{forw}}]$, with $V_{\mathrm{refl/forw}}$ being the amplitude and $\phi_{\mathrm{refl/forw}}$ the phase of the reflected and forward traveling wave, respectively. The rf data recorded in the data acquisition system [19] is manually recalibrated according to Ref. [20] in order to remove cross-coupling effects between the forward and reflected signals. The energy at the entrance of the module is calculated with phase and amplitude of the gun rf, $[V_{gun}, \phi_{gun}]$, assuming the gun calibration to accelerate the beam to 5.6 MeV/c at $V_{gun} = 54$ MV m$^{-1}$ and $\phi_{gun} = -1.85°$. In order to verify the model function, an experimental setup is designed in which the impact of unknown parameters, e.g., the offsets of BPMs, is supposed to cancel out. This is done by comparing different sets of rf parameters and BPM readings, as will be described later. Each measurement of BPM- and rf data is an average over approximately 300 consecutive bunch-trains to deal with short-term jitter. The bunch spacing is 1 μs.

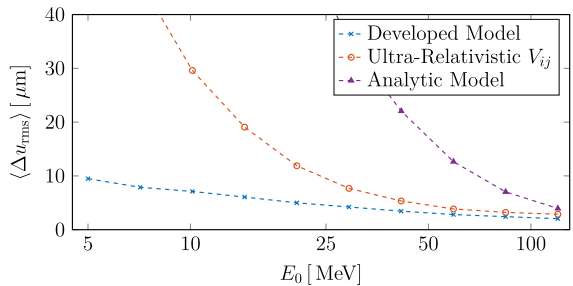

FIG. 5. Rms difference between the output of the model function and an ASTRA tracking $\langle\Delta u_{\mathrm{rms}}\rangle$ as a function of initial beam energy $E_0$, evaluated for one cavity and averaged over both transverse planes. The beam transport matrices are calculated with different models. Plotted is the developed model (blue, Eq. (13)), numerical transfer matrices with ultra-relativistic $V_{ij}$ (red, cf. [18]) and the analytic model (purple, cf. [12]).

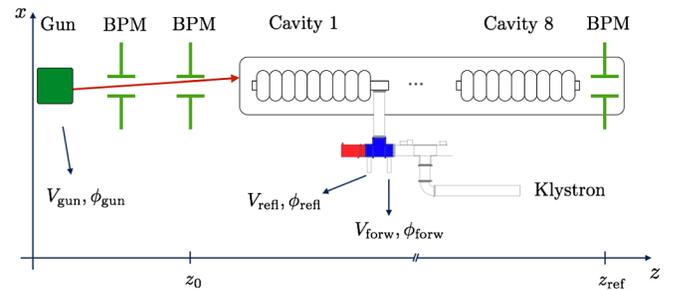

FIG. 6. Schematic drawing of the experimental setup at the injector section at FLASH. Two BPMs are located in the drift space between the gun and the first cavity. The last BPM is located inside the cryogenic vessel and gives the reference offset. The Klystron power is distributed in waveguides (white), circulators (blue), and dumped in loads (red). Amplitudes and phases of the forward and reflected waves are measured for all cavities at the circulators.

If the accelerating gradient is changed, the transverse off-axis fields change as well as the on-axis accelerating field of the cavity. If the variation of the klystron power is slow enough to ensure a steady-state condition, coupler kicks should vary only in strength, not in direction and in inferior order. A reference is measured. The klystron power is modulated with $f_{mod} = 3$ kHz, assuring the accelerating field to be in resonance. The difference of BPM readouts in respect to the reference measurement for the horizontal and vertical plane, $\Delta x$ and $\Delta y$, respectively, is plotted in Fig. 7 in black. The output of the model function is calculated and subtracted correspondingly and plotted in blue and red in Fig. 7. This, so to speak, partial derivative of the offset on the vector sum of the accelerating field shows a reasonable agreement with the BPM difference reading, considering that misalignments are not taken into account in the calculation.

As a second step, the implementation of coupler kicks is focused on. Caused by the limited bandwidth of the cavity, an increase of the modulation frequency of the klystron power will lead to a smaller amplitude of the modulation of the accelerating field, thus a higher reflected power. The ratio of the forward and reflected wave should vary and therefore, as pointed out in Fig. 4, the mainly horizontal forces induced by the power input coupler. Thus the impact of the power coupler compared to the overall transverse dynamics should increase with higher modulation frequencies. Especially misalignments should, for the most part, cancel out.

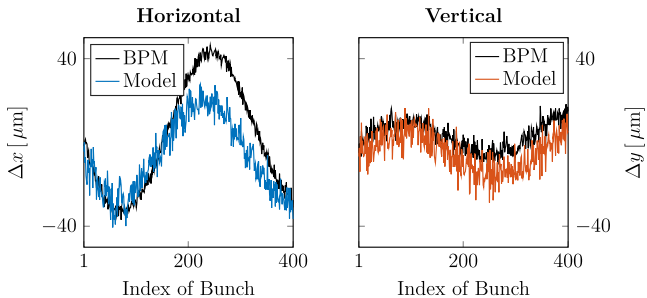

FIG. 7. Difference between the reference offset and the offset while a 3 kHz modulation is applied on the forward power. The BPM readout differences (black) and the corresponding model evaluations (colored) are plotted for the horizontal (left) and vertical (right) plane.

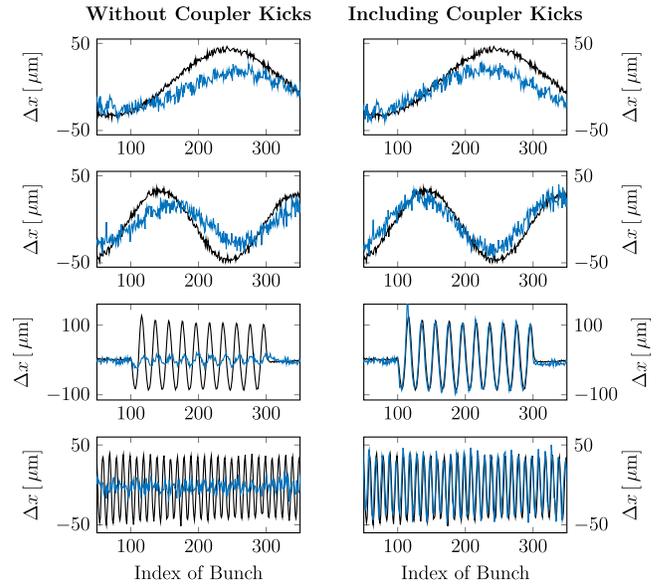

FIG. 8. Difference between the reference trajectory and the trajectory while applying modulations with different frequencies on the klystron power (from top to bottom: 3 kHz, 5 kHz, 50 kHz and 100 kHz). The BPM readout differences (black) and the corresponding model evaluations (blue) are plotted for the horizontal plane downstream the injector module at FLASH. The beam transport calculations are done both, excluding (left) and including (right) coupler kicks. At high modulation frequencies the transverse dynamic is dominated by the variation of coupler kicks.

Figure 8 shows the results of four measurements. The modulation frequency of the klystron power is increased subsequently from 3 kHz to 5 kHz, 50 kHz, and 100 kHz. Shown are the differences of the horizontal BPM readout between the modulated setup and the reference setup and the difference of the corresponding output of the model function in black and blue, respectively. The left column is calculated without coupler kicks, the right column shows the calculation including coupler kicks. Comparison between the columns in the last two rows points out that the beam dynamics above a modulation frequency of several 10 kHz are dominated by coupler kicks. These transverse dynamics are well described by the developed model.

It can be concluded that the implemented model function is both qualitatively and quantitatively able to reproduce the experimentally generated transverse trajectory features at the injector module.

## IV. DETERMINATION OF MISALIGNMENTS FROM MULTIBUNCH DATA

Due to the short computation time of the derived model function, numerical studies on the beam dynamics of long bunch-trains can be performed efficiently. In this section the deduction of misalignments from experimentally derived multibunch rf and BPM data is considered. It will be shown that the numbers of degrees of freedom, the limited amount of available data and the considered errors of the BPMs prevent reasonable fitting of the cavity and module misalignments.

Misalignments of involved structures are modeled by coordinate system transformations. The nomenclature used in this work is as follows: the beam parameters are $\Delta u$ for the trajectory offset and $\Delta u'$ for trajectory tilt angle with respect to the design axis, where $u$ stands for the transverse planes $x$ and $y$. The offset of a structure, for example the cavity or the module, is defined at its longitudinal center. The tilt of a cavity is evaluated around its longitudinal center, whereas the tilt of the module is evaluated around its entrance. The reference axis of the cavities is the axis of the module, while its reference is the design axis of the gun section. The misalignments of cavities and modules are $\Delta u_{\text{cav/mod}}$ and $\Delta u'_{\text{cav/mod}}$ for the offset and tilt angle, respectively. An algorithm was implemented, that allows simultaneous fitting of these parameters to an arbitrary amount of experimental data sets. This allows to increase the amount of experimental data, including the possibility that a certain set of rf and beam input parameters is insensitive to a particular misalignment, see Ref. [21] for more details.

Dedicated experiments [22] at the injector module at FLASH with 400 bunches are made to increase the sensitivity of single data sets to misalignments. The detuning of individual cavities and their loaded quality factors are subsequently changed. This results in a distinct change of rf parameters. Additionally, the beam input trajectory is varied. A total amount of 28 data sets are used for fitting. Each data set is averaged over 300 consecutive bunch-trains to deal with short-term jitter.

Despite the variation of rf parameters it was not possible to change the sensitivity of the model function with respect to misalignments of downstream cavities significantly. Apart from this, the amount of fit parameters, thus degrees of freedom, is large compared to the observable parameters. The consequential under-determination and the limitation of the fitting algorithm is evaluated first. Rf and beam input parameters from the above described data sets are used as input of the model function. Random misalignments are created. The output of the model function at the end of the module reflects a pseudo BPM reading and is used as reference for the fitting routine. Thus, it can be tested if the fit algorithm is able to identify the correct values of the misalignments.

In a first simplified example only cavity misalignments are studied. BPMs are supposed to work accurately and the module is on axis. A total set of $10^4$ random misalignments according to the specification limits of $\Delta u_{\text{cav}} = 0.5$ mm and $\Delta u'_{\text{cav}} = 0.5$ mrad are evaluated. The initial values of the fitting routine are created randomly within the upper and lower bounds $\Delta u_{\text{bound}} = \pm 0.5$ mm and $\Delta u'_{\text{bound}} = \pm 0.5$ mrad, respectively. The rms difference between the simulated misalignments and the fitted misalignments is shown in Fig. 9 for the horizontal plane. The difference is significant. Note that the rms value of a real number which is randomly distributed between $\pm 0.5$ is about 0.3. It can therefore be concluded that based on the experimental data available it is not possible to fit the misalignments of individual cavities to a reasonable accuracy.

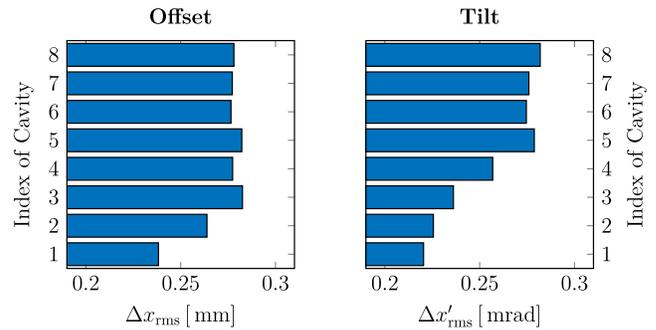

FIG. 9. Rms error between the fitted and the simulated misalignments of 8 cavities in the injector module. 28 experimentally derived rf data sets are used. No module misalignments or BPM errors are considered. The parameter range of the misalignments is set accordingly to the design values $\Delta u_{\text{cav}} = 0.5$ mm and $\Delta u'_{\text{cav}} = 0.5$ mrad.

As a second step the prediction accuracy of the fit algorithm regarding the module misalignment is estimated. A total set of $10^4$ random misalignments according to the uncertainties of $\Delta u_{\text{mod}} = 5$ mm and $\Delta u'_{\text{mod}} = 0.5$ mrad are evaluated. Analogously to the above mentioned simulation- and fit-procedure, the rms difference between the fitted and the simulated misalignments is calculated. Cavity misalignments and BPM errors are not considered. The rms error is found to be $\Delta u_{\text{mod,rms}} = 0.51$ mm and $\Delta u'_{\text{mod,rms}} = 0.29$ mrad.

However, errors of the calibration factor $\delta u_{\text{BPM}}$ and offset $\Delta u_{\text{BPM}}$ of the involved BPMs can not be neglected. The prediction accuracy of the fit algorithm is expected to depend significantly on the accuracy of the BPMs. Its particular value is estimated as follows. The misalignment of the module is fitted for different bounds of the BPM error factor $\Sigma$. It incorporates the error of the calibration factor and the offset of each BPM and will be defined such that $\Sigma = n$ corresponds to $\delta u_{\text{BPM}} = n \cdot 30\%$ and $\Delta u_{\text{BPM}} = n \cdot 0.5$ mm. For example, $\Sigma = 1$ reflects the possibility that all involved BPMs have a maximum calibration error of $\pm 30\%$ and maximum offset error of $\pm 0.5$ mm. The ratio between these values is reasonable for FLASH [23]. At each step $\Sigma_i$, the prediction accuracy of the fitted module misalignment is estimated analog to the above described method and used for classification of the upcoming fit.

Finally the unmodified recorded data sets are used for fitting. Fit parameters are the misalignment of cavities, the misalignment of the module in respect to the gun section and the calibration factors and offset errors of the BPMs. The boundary values for the misalignments are set according to the previously mentioned limits. The boundary values for the BPMs are incrementally increased according to the previously described ratio $\Sigma$. At each step $\Sigma_i$, 500 different sets of randomly created starting points for the fit algorithm are independently evaluated. Solutions with $\chi^2_{k,i} < 2 \cdot \chi^2_{\text{best},i}$ are accounted, where $\chi^2_{\text{best},i}$ describes the lowest found value for each $\Sigma_i$.

The results are shown in Fig. 10. The left-hand side shows the fitted module offset $\Delta u_{\text{mod}}$ in respect to the gun section as a function of the BPM error factor $\Sigma$. The horizontal and vertical plane are plotted in blue and red, respectively. The right-hand side shows the fitted values for the module tilt angle $\Delta u'_{\text{mod}}$. The crosses are fitted solutions, while the dashed lines indicate the estimated prediction accuracy of the fit algorithm based on the previously described simulations. In the lower row $\chi^2_{k,i}$ is plotted for each solution. The initial $\chi^2_0$ is in the order of $10^5$.

For perfectly calibrated BPMs, thus $\Sigma = 0$, the fitted offset spans a range of about 2 mm, whereas the range of the fitted tilt angle is 800 $\mu$rad. Note that the longitudinal size of the module is 12 m. A tilt angle of 800 $\mu$rad therefore corresponds to an offset of about 5 mm on both ends in respect to the design axis of the accelerator. The range of the fitted values is unreasonably large. Subsequent studies on the accuracy of the BPMs located in the injector section indicate that $\Sigma > 3$. Furthermore long term drifts, most likely caused by technical issues in the readout electronics were identified. The actual BPM calibration during the multibunch measurements can therefore not be determined beyond reasonable doubt. It can be concluded that the recorded data sets are not distinct and comprehensive enough to deal with the large amount of degrees of freedom and the parameter limits of the regarded system. If the module misalignment should be determined sufficiently by multibunch-based misalignment measurements, several tasks would have to be accomplished. At FLASH, a quadrupole-triplet is located between the first and the second BPM downstream from the injector module. The difference between the theoretical and actual transfer matrix of this magnet is known to be significant [24]. If this problem is solved, both downstream BPMs could be used for the analysis. This results in full phase space information at the reference plane. In addition, dedicated studies on the calibration of the involved BPMs are advised.

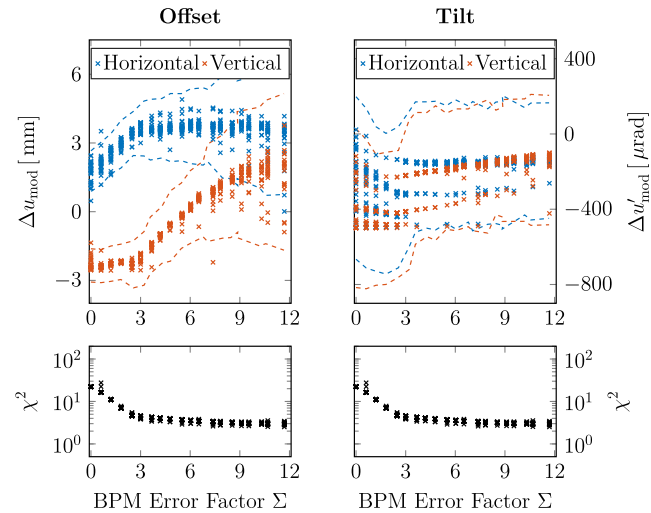

FIG. 10. Fitted offset (left) and tilt angle (right) of the injector module at FLASH for the horizontal (blue) and vertical (red) plane. The fit results (top) and the corresponding $\chi^2$ (bottom) are plotted as a function of the BPM error factor $\Sigma$. $\Sigma$ includes the calibration factor $\delta u_{\text{BPM}}$ and the error of the offset value $\Delta u_{\text{BPM}}$ in a fixed ratio. $\Sigma = n$ corresponds to both, $\delta u_{\text{BPM}} = n \cdot 30\%$ and $\delta u_{\text{BPM}} = n \cdot 0.5$ mm. The crosses are fitted solutions, while the dashed lines indicate the prediction accuracy based on simulations.

## V. STATISTICAL INVESTIGATION ON INTRABUNCH-TRAIN TRAJECTORY VARIATIONS

Statistical studies are made in order to generally describe the influence of misalignments in correlation with variable rf parameters on the intrabunch-train transverse dynamics. As a figure of merit it is reasonable to introduce the multibunch emittance blow-up

$$\tau = \frac{\epsilon_{\text{MB}}}{\epsilon_{\text{SB}}} \quad (16)$$

in order to describe the ratio of the multibunch emittance $\epsilon_{\text{MB}}$ in respect to the emittance of a single bunch $\epsilon_{\text{SB}}$. The multibunch emittance is calculated according to Refs. [25,26], assuming that only centroid dynamics differ within one bunch-train. The transverse single bunch emittance is assumed to be 1 $\mu$m and the Courant-Snyder parameters at the reference plane are assumed as $\alpha_x = -0.076$, $\alpha_y = -0.049$, $\beta_x = 12.15$ m, and $\beta_y = 12.18$ m. As described in this paper, the model requires the voltage standing wave ratio $\Gamma$ as input. However, its variability within one bunch-train is not a common specification. The following workflow was developed in order to generate artificial rf data sets, while obeying the limitations set by the actual low level rf setup. The amplitude $\bar{V}_0$ and phase $\bar{\phi}$ of the vector sum of the accelerating field have to be chosen as well as the maximum slope of the amplitude $\Delta V_0$ and phase $\Delta \phi$ of the accelerating field of the individual cavities and their maximum detuning $\Delta f$ within one bunch-train. The mean accelerating fields of the individual cavities are derived according to the waveguide-setup at the injector module at FLASH. The difference of the individual mean amplitude to the one eighth of the vector sum is calculated. The individual cavity accelerating field slope is randomly created, while keeping the vector sum constant. For each cavity, the signals of the forward and reflected wave have to be calculated. A detailed discussion on rf cavities can be found in Ref. [7]. We assume the special case of a superconducting cavity operating close to the steady state condition, nearly on crest, with beam loading and a detuning small compared to the resonance frequency (see Sec. 3.3.2 in Ref. [7]). The cavity voltage $\mathbf{V}_{\text{cav}}$, the forward- and reflected wave $\mathbf{V}_{\text{for}}$ and $\mathbf{V}_{\text{ref}}$ and the beam current $\mathbf{I}_b$ are then related according to

$$\mathbf{V}_{\text{cav}} = \frac{1 + i\tan\psi}{1 + \tan^2\psi} \cdot \left[2\mathbf{V}_{\text{for}} + \frac{1}{2}\left(\frac{R}{Q_0}\right)Q_L \cdot \mathbf{I}_b\right]$$
$$\mathbf{V}_{\text{cav}} = \mathbf{V}_{\text{for}} + \mathbf{V}_{\text{ref}}, \qquad (17)$$

where the bold letters indicate complex numbers, for example $\mathbf{V}_{\text{ref}} = V_{\text{ref}} \cdot e^{i\phi_{\text{ref}}}$. The ratio of the shunt impedance $R$ and the unloaded quality factor $Q_0$ depends only on the geometry of a cavity, where the loaded quality factor $Q_L$ in this special case depends only on the coupling between the cavity and the waveguide system. The detuning angle $\psi$ is defined as

$$\tan\psi = 2Q_L \frac{f_0 - f}{f_0} = 2Q_L \frac{\Delta f}{f_0} \qquad (18)$$

with $f_0$ being the resonance frequency of the cavity and $f$ the operating frequency, leading to the detuning $\Delta f = f_0 - f$. Using Eqs. (18) and (17) and assuming a constant charge and repetition rate of the bunches, the forward and reflected wave can be expressed as a function of the phase difference $\phi_{\text{cav}}$ between the bunch and the cavity voltage with an amplitude of $V_{\text{cav}}$ and the detuning $\Delta f$. The voltage standing wave ratio $\Gamma$ follows. With these rf parameters the transfer matrices and coupler kick coefficients of the model function are calculated for each bunch and each cavity individually.

In order to avoid the loss of generality by analyzing only a coincidental case of parameters, a Monte Carlo filtering based method, also referred to as regional sensitivity analysis [27], is being followed. The goal is to analyze a multidimensional stochastic output statistically by conditioning the input space. The fundamental idea should be outlined on the basis of a simple example first. Consider the value of interest $\tau$ is a function of several parameters $x$, $y$. Consider the projection of $\tau$ on the $\tau$–$x$-plane to be distributed as shown in Fig. 11 in the upper row on the left. In the upper right its correspondent cumulative distribution function (CDF) of $\tau(x)$ is drawn. The CDF of $\tau(x)$ accumulates the probability that the related variable takes values less than or equal to the evaluation point of the CDF. The interpretation of $\text{CDF}(\tau(x) = 2) = 0.5$ is therefore, that half of the evaluations of $\tau(x)$ have a value smaller than or equal to 2. For the upcoming analysis a critical value $\tau_c$ will be defined, such that $\text{CDF}(\tau_c) = p_c$. A reasonable value is $p_c = 0.9$, hence 90% of the evaluations of $\tau$ have values smaller than or equal to $\tau_c$. Note that in the first example of Fig. 11 the particular value of $x$ at which $\tau(x)$ is evaluated has no impact on the distribution of $\tau$. Thus $\tau_c$ is not sensitive to the range of $x$. Now consider the projection of $\tau$ on the $\tau$–$y$-plane to be distributed as shown in the mid row of Fig. 11. Obviously the distribution of $\tau$ depends on the range of $y$. In this example, however, the corresponding CDF is calculated unconditionally, meaning that the whole parameter range of $y$ is considered. The resulting CDF is identical to the one shown in the upper example. In the third example in the lower row, conditions on the input space are applied. The range of $y$ is binned into two subranges, $y_1$ and $y_2$, as highlighted with different colors in the lower row. The conditional CDF of $\tau(y)$ depends on the range of the $y_i$. Therefore the critical values of the two conditional CDFs, $\tau_{c,1}$ and $\tau_{c,2}$, differ from each other. The sensitivity of the model to its parameters can be quantified in the variance of the critical values $\text{var}(\tau_{c,i})$. It indicates if a model parameter, in this example $y$, is influential in determining the distribution of the model output $\tau$ (cf., the upper and lower row of Fig. 11). It is worth noting that the particular distribution of $\tau_{c,i}$ depends on the parameter limits, the size of the subsamples and on the definition of $p_c$.

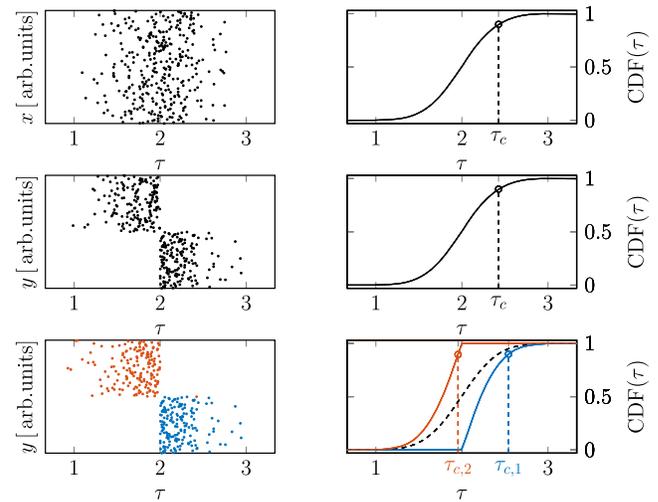

FIG. 11. Qualitative illustration of the difference between a conditional and unconditional cumulative distribution function (CDF). The upper row shows the CDF (left) of a normally distributed variable $\tau(x)$. The value of $x$ has no impact on the distribution of $\tau$. The distribution of $\tau(y)$ (mid row) differs from $\tau(x)$, while the unconditional CDF remains unchanged. The lower row shows two conditional CDFs (blue and red) of $\tau(y)$ with the correspondingly colored input samples of $y$.

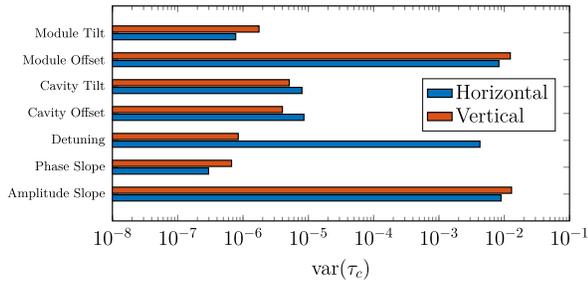

FIG. 12. Single-parameter variance of $\tau_c$, calculated for 5 subsamples each, for the horizontal (blue) and vertical plane (red). Amplitude slope of the accelerating field and the offset of the module clearly dominate the variation of $\tau_c$. The detuning only impacts the horizontal plane. Note that the plotted results strongly depend on the defined parameter limits.

The previous concept will be applied on the analysis of the intrabunch-train transverse dynamics in the injector module. In this case the different parameters are: the maximum values of the intrabunch-train variation of the amplitude $\Delta V_0$ and phase $\Delta\phi$ of the accelerating field, respectively, and the variation of the detuning $\Delta f$. In addition, cavity offsets $\Delta u_{\text{cav}}$ and -tilts $\Delta u'_{\text{cav}}$ in respect to the module design axis as well as the offset of the module $\Delta u_{\text{mod}}$ and its tilt $\Delta u'_{\text{mod}}$ in respect to the design axis of the gun section are studied. In the following, $\Delta V_0$ and $\Delta\phi$ will be referred to as amplitude slope and phase slope, respectively.

The total number of model evaluations is $10^6$. The parameters are varied randomly within the range shown in Table I in the Appendix, giving a seven-dimensional distribution of $\tau$-values. When conditions are applied subsequently to each parameter, the variation of $\tau_c$ provides information about the distribution of $\tau$ in respect to that parameter. Figure 12 points out this one dimensional relationship for $p_c = 0.9$. For each parameter the sample size is divided into 5 subsamples. For example, the conditional limits on the offset of the injector module varies between $\Delta u_{\text{mod},1} = (0.5 \pm 0.5)$ mm, $\Delta u_{\text{mod},2} = (1.5 \pm 0.5)$ mm, and so on. The determination of the multi-bunch emittance blowup clearly is dominated by the amplitude slope, the detuning (for the horizontal plane) and the offset of the injector module with respect to the gun section, whereas other parameters are not significant in determining the distribution of $\tau$. Results in Fig. 12 have to be taken cautiously. The range of each parameter is binned into relatively large subranges and single-parameter analysis does not reveal correlations between individual model parameter. Note also that the parameter limit for the module offset is ten times larger than the limit of individual cavity offsets. However, since the individual parameter range was set according to reasonable tolerance limits, this approach clearly indicates the dissimilar influence of parameters and assists in finding key parameters for the upcoming analysis.

For a more precise prediction the size of the subsamples has to be decreased, while conditions can be applied to multiple parameters simultaneously. Thus, analysis of $\tau_c$ can be used to point out $n$-dimensional correlations. At first a parameter prioritization follows. The phase slopes of the accelerating field of individual cavities $\Delta\phi$ as well as their offsets $\Delta u_{\text{cav}}$ and tilts $\Delta u'_{\text{cav}}$ are secondary in limiting the multibunch emittance for the current parameter range. Higher order mode based cavity misalignment measurements indicate, that the intended misalignment tolerances for individual cavities have been met throughout the modules, thus providing fixed evaluation limits of $\Delta u_{\text{cav}} = 0.5$ mm and $\Delta u'_{\text{cav}} = 0.5$ mrad. The maximum module tilt angle will be set to $\Delta u'_{\text{mod}} = 0.5$ mrad. The maximum slope of the phase of the accelerating field will be set to $\Delta\phi = 1°$. Experimental observations show that this value is hardly exceeded during a typical user run with long bunch-trains. Fixing the above mentioned values remains a three-dimensional parametrization of $\tau_c(\Delta V_0, \Delta f, \Delta u_{\text{mod}})$, representing most of the variance of the multibunch emittance for realistic scenarios. This representation allows to study the three-parameter-CDF on $\tau$, of which a two-dimensional projection is plotted in Fig. 13 as a contour plot of $\tau_c(\Delta V_0, \Delta f)$ for different offset limits of the injector module. The number of subsamples for each parameter is ten, splitting the total amount of $10^6$ evaluations for each module offset into $10 \times 10 = 100$ subsamples, each containing about $10^4$ data points.

The superior influence of the amplitude slope compared to the detuning for the vertical plane is obvious. However, a critical amount of multibunch emittance is only exceeded when both amplitude slope and the module offset reach large values simultaneously. It is reasonable to assume that the evaluation limits are chosen large enough to include all possible machine realizations. The results can therefore be used in order to quantify the achievable performance improvement when reducing one of the parameter limits. If the goal, for instance, is to assure by a likeliness of 90%, that the horizontal multibunch emittance blowup is below 30%, one has to limit either the amplitude slope of the accelerating field to 0.7 MV m$^{-1}$ or fix the module offset in respect to the gun section to better than 2 mm. Limiting the detuning, on the other hand, for example by means of a piezo-tuner to 10 Hz [28], would decrease $\tau_c$ by about 15%, depending on the remaining module offset and amplitude slope.

Analysis of experimental data at FLASH [21] shows intrabunch-train trajectory variations up to 200 $\mu$m being induced in the injector module in both planes. This reflects a multibunch emittance blowup of about 20%. The rms-value of the amplitude slope of the accelerating field is about 250 kV m$^{-1}$ with a detuning of about 100 Hz. The presented simulations are therefore interpreted as an indicator for a significant offset of the injector module with respect to the gun section, especially in the vertical

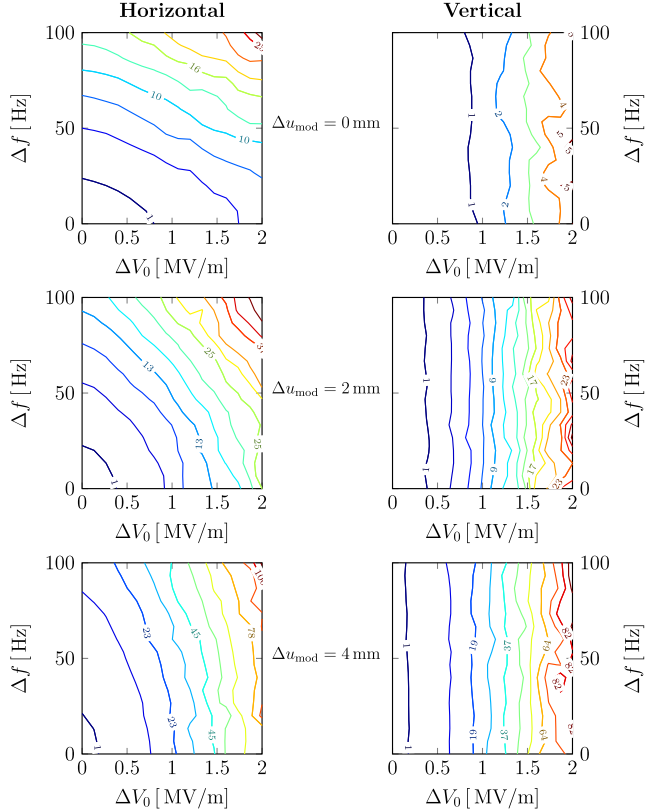

FIG. 13. Analysis of the two-dimensional conditional CDF of the multibunch emittance blowup $\tau$ as a function of detuning, $\Delta f$, and amplitude slope of the accelerating field, $\Delta V_0$, for different offsets of the injector module $\Delta u_{\mathrm{mod}}$. The contour lines of the critical value $\tau_c$ for the horizontal (left column) and vertical (right column) plane are plotted in %, e.g., 23 corresponds to $\tau_c = 1.23$. Each plot contains $10^6$ evaluations which are binned into $10 \times 10 = 100$ subsamples for the calculation of the $\tau_{c,i}$.

plane. This result agrees with the cautious estimate from fitting as presented in Fig. 10.

## VI. SUMMARY AND CONCLUSION

An efficient model for calculating the beam transport in a TESLA cavity at low beam energy was found, including an analytical expression for describing discrete coupler kicks. Cross-checking between tracking and experimental data was performed. It can be concluded that the presented model is both qualitatively and quantitatively able to reproduce the transverse beam dynamics at low beam energy reasonably well.

The deduction of misalignments of accelerating structures from multibunch rf- and BPM data was considered and a fit algorithm was implemented. Thus far experimental constraints prevent accurate fitting of the misalignments of cavities and the injector module. On that account the developed model was used to analyze the interactions of intrabunch-train rf variations and structure misalignments statistically. Results indicate that the observed intra-bunch-train trajectory variation at the injector section at FLASH is caused by significant misalignments of the injector module with respect to the gun section, which is supported by the preliminary estimates of the fitting results. Investigations on procedures to reduce the misalignment are advised. It was furthermore possible to give accurate bounds for a performance study of the injector module regarding intra-bunch-train trajectory variations. It can be inferred that limiting the amplitude slope of the accelerating field is crucial for an improvement of the multibunch performance. Additionally Lorentz force detuning compensation would decrease the horizontal multibunch emittance blowup by 15% solely by limiting the variation of coupler kicks.

## ACKNOWLEDGMENTS

This research was supported by the German Academic Scholarship Foundation. We would like to thank Vladimir Balandin, Siegfried Schreiber, Nicoleta Baboi, Jörg Rossbach and Christian Schmidt for valuable comments and fruitful discussions. We wish to extend our particular thanks and appreciation to Sven Pfeiffer, who with great skill and patience supported the measurements.

## APPENDIX: SIMULATION PARAMETERS

TABLE I. Parameter range for simulations at the injector module. Listed are the maximum values. Variations are considered to be within one bunch-train.

| Parameter | Value | Description |
|---|---|---|
| $n_{\mathrm{bunch}}$ | 400 | Number of Bunches per Bunch-Train |
| $E_0$ | 5.6 MeV | Initial Beam Energy |
| $\overline{u_0}$ | 0 mm | Mean Initial Offset |
| $\overline{u_0'}$ | 0 mrad | Mean Initial Angle |
| $\Delta u_0$ | 0 mm | Variation of Initial Offset |
| $\Delta u_0'$ | 0 mrad | Variation of Initial Angle |
| $\overline{V_0}$ | 155 MV m$^{-1}$ | Vector Sum of Acc. Field Amplitude |
| $\bar{\phi}$ | $-5°$ | Vector Sum of Acc. Field Phase |
| $\Delta V_0$ | 2 MV m$^{-1}$ | Var. of Ind. Cavity Acc. Field Amplitude |
| $\Delta \phi$ | 4° | Var. of Ind. Cavity Acc. Field Phase |
| $\Delta f$ | 100 Hz | Var. of Ind. Cavity Detuning |
| $\Delta u_{\mathrm{cav}}$ | 0.5 mm | Offset of Cavities |
| $\Delta u_{\mathrm{cav}}'$ | 0.5 mrad | Tilt of Cavities |
| $\Delta u_{\mathrm{mod}}$ | 5 mm | Offset of Module |
| $\Delta u_{\mathrm{mod}}'$ | 0.5 mrad | Tilt of Module |